\pdfoutput=1
\documentclass{acm_proc_article-sp}
\usepackage{graphicx}
\usepackage{subfigure}
\title{@I to @Me: \\ An Anatomy of Username Changing Behavior on Twitter}
\author{Paridhi Jain, Ponnurangam Kumaraguru \\ Indraprastha Institute of Information Technology (IIIT-Delhi), India \\ \{paridhij, pk\}@iiitd.ac.in \\ {precog.iiitd.edu.in}}

\begin{document}
\maketitle
\begin{abstract}

An identity of a user on an online social network (OSN) is defined by her profile, content and network attributes. OSNs allow users to change their online attributes with time, to reflect changes in their real-life. Temporal changes in users' content and network attributes have been well studied in literature, however little research has explored temporal changes in profile attributes of online users. This work makes the first attempt to study changes to a unique profile attribute of a user -- \emph{username} and on a popular OSN which allows users to change usernames multiple times -- \emph{Twitter}. We collect, monitor and analyze 8.7 million Twitter users at macroscopic level and 10,000 users at microscopic level to understand username changing behavior. We find that around 10\% of monitored Twitter users opt to change usernames for possible reasons such as space gain, followers gain, and username promotion. Few users switch back to any of their past usernames, however prefer recently dropped usernames to switch back to. Users who change usernames are more active and popular than users who don't. In-degree, activity and account creation year of users are weakly correlated with their frequency of username change. We believe that past usernames of a user and their associated benefits inferred from the past, can help Twitter to suggest its users a set of suitable usernames to change to. Past usernames may also help in other applications such as searching and linking multiple OSN accounts of a user and correlating multiple Twitter profiles to a single user.


\end{abstract}

\category{H.3.5}{Online Information Services}Online Web Services {}
\section{Introduction} \label{Introduction}
\begin{figure*}[t]
\centering
   \subfigure[User-ID `12x6917x09' recorded on November 8, 2013, holding the username `alone\_trix!' ]{
   \includegraphics[scale=0.27]{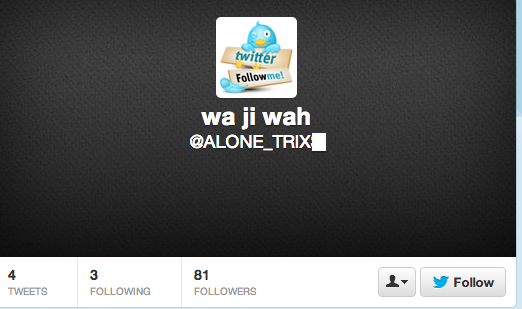}\label{fig:change1}
   }
   \quad
   \subfigure[User-ID `12x6917x09' recorded on January 15, 2014, changed her username to `JOX\_4!' ]{
   \includegraphics[scale=0.27]{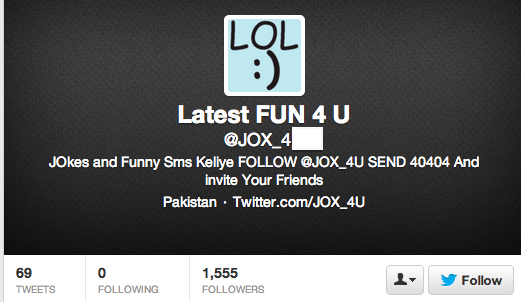}\label{fig:change2}
   }
   \quad
   \subfigure[User-ID `55x814x82' recorded on January 15, 2014, holding the username `alone\_trix!']{
   \includegraphics[scale=0.27]{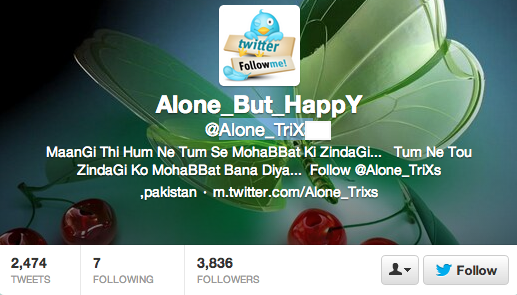}\label{fig:change3}
   }
   \caption{ shows a Twitter user who changes her username over time (a, b). If the user ID is not known, her new username cannot be found. Two different users with two different unique IDs pick the same username `alone\_trix!' at two different timestamps (a, c). A Twitter search for user ID `12x6917x09' with her old username `alone\_trix!' redirects to a different user with user ID `55x814x82'. }
      \label{fig:change}
\end{figure*}

Since 2009, \textit{Twitter} has become a popular online social network (OSN) among millions of users. As of 2013, Twitter has 200 million users creating more than 400 million tweets per day, making it the most populous micro-blogging service.~\footnote{https://blog.twitter.com/2013/celebrating-twitter7} Users join Twitter via an easy sign-up process. During the sign up process, users set some profile attributes as part of their account, where some are mandatory (`username') and some optional (e.g. name, location). \textit{Username} attribute is an important attribute, with which a user can be referred, searched and tagged \emph{uniquely} in a tweet by any other user on Twitter. After the sign up process is complete, Twitter assigns a \emph{unique}, numeric, and constant \emph{ID} to the user, which can only be known to other users via Twitter's API request, and not via Twitter web / mobile interface. A user is therefore associated with two unique attributes on Twitter -- username and user ID.  \\
\indent User ID is not changeable however, changes to username are allowed according to user convenience and requirements.~\footnote{https://support.twitter.com/articles/14609-changing-your-username} Figure~\ref{fig:change} shows a Twitter user who changes her username from `alone\_trix!' to `JOX\_4!'. Few OSNs such as Facebook~\footnote{https://www.facebook.com/help/105399436216001\#What-are-the-guidelines-around-creating-a-custom-username?} put a sealing on the number of times a user can change her username, while Twitter does not. Allowing changes to a username is beneficial because it may help a user to accommodate her changing attributes, likes, dislikes, and her changing interests over time.  However, such username changes may lead to unwanted consequences -- a  user search with her past username and with no information of her numeric user ID, may lead to non-searchability (no results) or unreachability (broken link~\footnote{Sorry, the page doesn't exist page on Twitter}) to the user's profile. Further, Figure~\ref{fig:change} shows a scenario where search for a Twitter user, who had username `alone\_trix!' at an earlier timestamp, redirects to a different user who picked the same username at a later timestamp. If changing usernames may lead to loss of connectivity to a user profile, broken links, or redirection to a  different user profile, no user may intend to do it. However, we observe that around 10\% of observed 8.7 million Twitter users change their usernames over time (as discussed in Section~\ref{Methodology}). \\
\indent To the best of our knowledge, there is very little work to understand why users temporally change profile attributes on OSNs, specifically username. We make the first attempt to explore this behavior of changing usernames temporally and term it as \emph{username changing behavior}, in this paper. We try to understand properties of users who temporally change usernames and explore a set of reasons which explain why users change usernames over time and what usernames they switch to. Our contributions are: 
\begin{itemize}
\item
\vspace{-3mm}
We present the first longitudinal study to understand username changing behavior on Twitter. We find that around 10\% of 8.7 million Twitter users change their usernames temporally.
\item
We analyze a focused set of users exhibiting username changing behavior and observe that few users change their usernames frequently, and within 24 hours of the earlier username change. Users choose less similar new usernames while a few favor one of their past usernames as their new username. 
\item
We show that frequency of changing username is weakly correlated with users' in-degree, activity and year of account creation. Users who change usernames are more popular, and active on Twitter than users who do not change.
\item
We learn that users change their usernames for reasons such as to gain space in a tweet, to gain more followers, to promote usernames, to suit a trending event, to gain / lose anonymity or to avoid boredom. 
\end{itemize}
\vspace{-3mm}
\indent We believe that a comprehension of username changing behavior can help Twitter to develop tailored username suggestion algorithms for its users. Currently, Twitter offers username suggestions only during sign up process and not later. We think that Twitter can suggest suitable usernames to a user, even after her account creation, based on her activity with her past usernames and associated profits / risks. Further, we think that a record of past usernames can be helpful for planning promotional content for users on Twitter, for correlating multiple Twitter profiles to a single real-world user, and for judging usernames of users on other OSNs (explored in Section~\ref{Applications}). \\ 
 \indent The paper is organized as follows. We first discuss the detailed methodology in Section~\ref{Methodology}. We then discuss username changing behavior characteristics in Section~\ref{In-Depth}, present properties of users who change usernames in Section~\ref{user} and then explore some plausible reasons for such behavior in Section~\ref{reasons}. We present some applications of our work in Section~\ref{Applications}. We then discuss the applicability and generalizability of our observations in Section~\ref{Discussion}, present related work and conclude the paper with some future directions.

\section{Methodology} \label{Methodology}
To qualitatively and quantitatively study behavior of changing username on Twitter, we collect 8.7 million users on Twitter and monitor them at regular intervals from October 1, 2013 - November 26, 2013. We also create a smaller set of 10,000 users who are selected to be monitored every 15 minutes for any username change. We describe our data collection framework now. 

\subsection{Data collection} \label{DataCollection}
Our data collection methodology is divided into two stages -- Seed collection, and Seed monitoring. Seed collection stage collects a set of users who are monitored in Seed monitoring stage. Details of each stage is as follows:
\subsubsection{Seed collection}
We collect a seed set of  8,767,576 users recorded by an event monitoring tool, MultiOSN~\cite{Dewan:2013:MRM:2528228.2528235}. Researchers shared the profiles of users who tweeted at least once about any of the 17 events (see Table~\ref{list_events}) monitored by MultiOSN, during April 1, 2013 - September 3, 2013. We refer to the seed set of 8,767,576 users as 8.7M users in the rest of the paper.
\begin{table}[ht]
\begin{center}
\small
\begin{tabular}{|p{4.5cm}|p{3.1cm}|}
\hline
{\bf{Global Events}} & {\bf \raggedright{Local Events} }  \\ \hline
Indian Premier League (IPL) & Bangalore Blasts \\ \hline  
Boston Blasts &  Uttrakhand Floods \\ \hline
Texas Fertilizer Plant Blast & Bodhgaya Blasts \\ \hline
Oklahoma Tornado & Telangana State \\ \hline
Champions Trophy Cricket & FoodBill \\ \hline
Nelson Mandela in ICU & Onion Crisis \\ \hline
Royal Baby & AsaramBapu Conviction \\ \hline 
\raggedright{Earthquake\_Pakistan ($ \rm{16^{th}}$ April)}  &  \\ \hline
\raggedright{Earthquake\_MiddleEast ($ \rm{1^{st}}$ May)}  &  \\ \hline
Mother's Day Parade shooting &  \\ \hline

\end{tabular}
\caption{Events monitored to collect seed users.}
\label{list_events}
\end{center}
\vspace{-8mm}
\end{table} 

\subsubsection{Seed monitoring}
In this stage, we monitor temporal changes in profile attributes of 8.7M user profiles, collected in the earlier stage. We query 8.7M users four times after definite intervals during October 1, 2013 - November 26, 2013, via Twitter Search API~\footnote{https://dev.twitter.com} and record their observed profile attributes in a database along with a timestamp. The process of querying 8.7M users is termed as \textit{scan}, in this paper. Table~\ref{collection} describes the time duration of each scan. Due to varying Internet speeds at servers and use of less number of authentication tokens during initial querying, each scan took different number of days to record 8.7M user profiles. Further, not all user accounts are activated during each scan, some deactivate / delete their accounts, while some are suspended by Twitter. We, therefore, record different number of user profiles in each scan.  \\
\indent We analyze four scans of 8.7M users collected over a period of 2 months (Oct 1 - Nov 26). A user is marked to have changed her username, if her username values in two consecutively timed scans are different. By comparing two consecutive scans, old and new usernames of a user are recorded. Note that, Twitter usernames are case-insensitive, therefore any case changes are not counted as username changes. We find that 853,827 users of 8.7M users (10\%) change their usernames at least once during our observation period (Oct 1 - Nov 26). Further, 853,827 users constitute 904,518 username change instances implying that a few users change their usernames multiple times. Therefore, we believe that it is significant to explore the characteristics of username changing behavior on Twitter. However, four scans of 8.7M users lack in necessary data. \\ \indent Due to huge number of users to query and limited Twitter API calls, each scan took long time to query each user in 8.7M dataset. Due to long scan stretches and intermediate intervals, we could neither record the exact date and time when users actually changed usernames nor all username changes a user went through. To capture the actual time and date when users changed their usernames as well as capture most username change instances triggered by users, we needed to scan 8.7M users at short intervals. Given Twitter allows 60 API calls per 15 minutes for users/lookup,~\footnote{https://dev.twitter.com/docs/rate-limiting/1.1/limits} scanning 8.7M users at short intervals (every 15 minutes) would require 1,462 application authentication tokens. We, therefore, use limited authentication tokens to respect the Twitter API resources utilization and initiate a fifteen-minute scan.
\subsection{Fifteen-minute scan}
Our focus in this research is on the users who change their usernames over time. We, therefore, first select 711,609 users who change their usernames at least once, during October 1, 2013 - November 15, 2013, as recorded during our Scan-I, Scan-II and Scan-III. Out of these 711,609 users, we \emph{randomly} sample 10,000 users~\footnote{We are continuously collecting data for the 711,609 users and hope to have a larger dataset soon.} and start to monitor them at short intervals. We query 10K users via Twitter API every 15 minutes. We term the faster scan of 10K users as \emph{Fifteen-minute scan}. Fifteen-minute scan starts on November 22, 2013; we bookmark the scan till March 19, 2014 and use this 117 days scan dataset for our analysis.~\footnote{We continue to scan 10K users after Mar 19, 2014 and record any username change.} If an observed username of a user profile exhibits a change in comparison to the most recent username recorded either in Scan-III or in fifteen-minute dataset which recorded her past username changes, fifteen-minute scan records the user profile in the fifteen-minute dataset and timestamp it. With fifteen-minute scan, we could record the exact timestamp when user changed her username, with an error limit of 15 minutes. Further, we observe that our regular Scan-IV took only one snapshot while fifteen-minute scan took 794 snapshots of 10K users, during Nov 22 - Nov 26. Scan-IV misses 712 username change instances triggered by 607 users, well captured by fifteen-minute scan. Therefore, fifteen-minute scan is successful in capturing most username changes made by the monitored users.

\begin{table}[h]
\begin{center}
\small
\begin{tabular}{|p{1.5cm}|p{2.3cm}|p{1.5cm}|p{1.5cm}|}
\hline
{\bf{Name of scan}} & {\bf \raggedright{Period of scan} }	& \raggedright{\textbf{\# users queried}} & {\textbf{\# users recorded} } \\ \hline
Seed set & Apr 1 - Sep 3 & 8,767,576 & 8,767,576 \\ \hline
Scan-I & Oct 1 - Oct 16 & 8,767,576 & 8,380,827  \\ \hline
Scan-II & Oct 25 - Oct 30 & 8,767,576 & 8,396,594  \\ \hline
Scan-III & Nov 8 - Nov 15 &  8,767,576 & 7,271,129 \\ \hline
Scan-IV & Nov 22 - Nov 26 & 8,767,576 & 8,388,010 \\ \hline
Fifteen-minute Scan & Nov 22 - Mar 19 & 10,000 & 2,698 \\ \hline
\end{tabular}
\caption{describes our four scans of 8.7M users and fifteen minute scan of 10K users.  }
\label{collection}
\end{center}
\end{table} 
\indent In the rest of the paper, we use the following definitions and notations. Each time a user changes her username, an event is counted and is termed as \emph{username change event / instance}. Old (dropped) username of a user $i$ at time $t_{1}$ is denoted by $u_{io}$ and new username at time $t_{2}$ is denoted by $u_{in}$, where $t_{1} < t_{2}$.  A user $i$ may use different usernames at different times, thereby creating multiple username change instances and constituting a \emph{username sequence}, denoted by $u_{i1}, u_{i2}, u_{i3}, \dots, u_{in}$. A user $i$ may choose her old usernames again, if still available. Such usernames are termed as \emph{revisited-usernames} and are denoted by $u_{ir}$. 

\subsection{Representativeness of the dataset} \label{Representativeness}
\begin{figure}[htbp]
   \centering
   \includegraphics[scale=0.47]{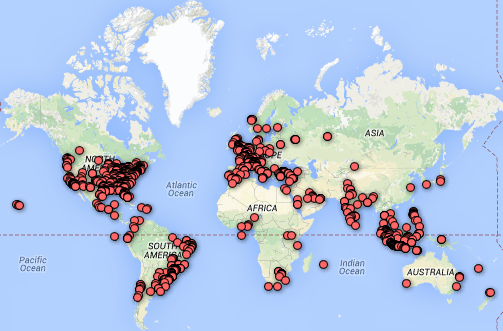} 
   \caption{shows geographical distribution of 926 users who geo-tagged their tweets. We see that users in our dataset span diverse geographical locations. }
   \label{fig:map}
\end{figure}

We examine geographical locations of 10K users to understand if they span across diverse locations. 
Geographical location of a user can be estimated by user's two profile attributes -- `location' and `timezone'. Location attribute of a user is an unformatted text, where user can provide any set of characters describing her location, such as `Moon'. Timezone attribute is selected from  a drop down list of timezones and user can decide to use any timezone. Both, location and timezone attribute may hold incorrect values, which may bias our understanding on representativeness of the dataset. We therefore use geo-tagged tweets by the users, to record their location. We map 1,849 unique latitude, longitude pairs from where 926 users (9\% of 10K) have posted their tweets (see Figure~\ref{fig:map}). We observe that users in our dataset, tweet from different locations around the world and not biased to only a few locations. Therefore, our analysis and results can be generalized to Twitter population from various global locations, who opt to change their usernames over time.

%
%

\begin{figure*}[htbp]
\centering
   \subfigure[]{
   \includegraphics[scale=0.15]{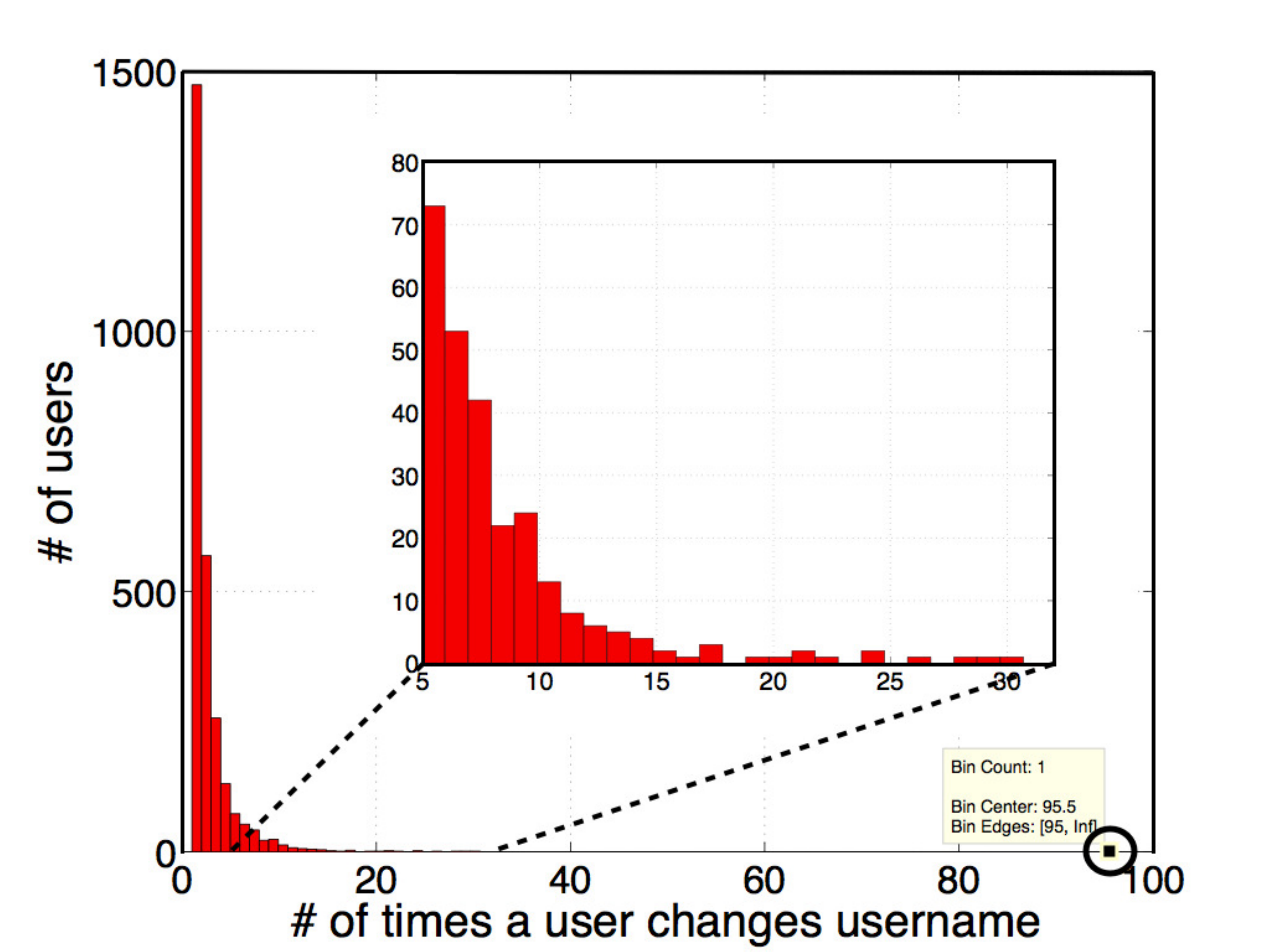}\label{fig:frequency1}
   }
   \quad
   \subfigure[]{
   \includegraphics[scale=0.23]{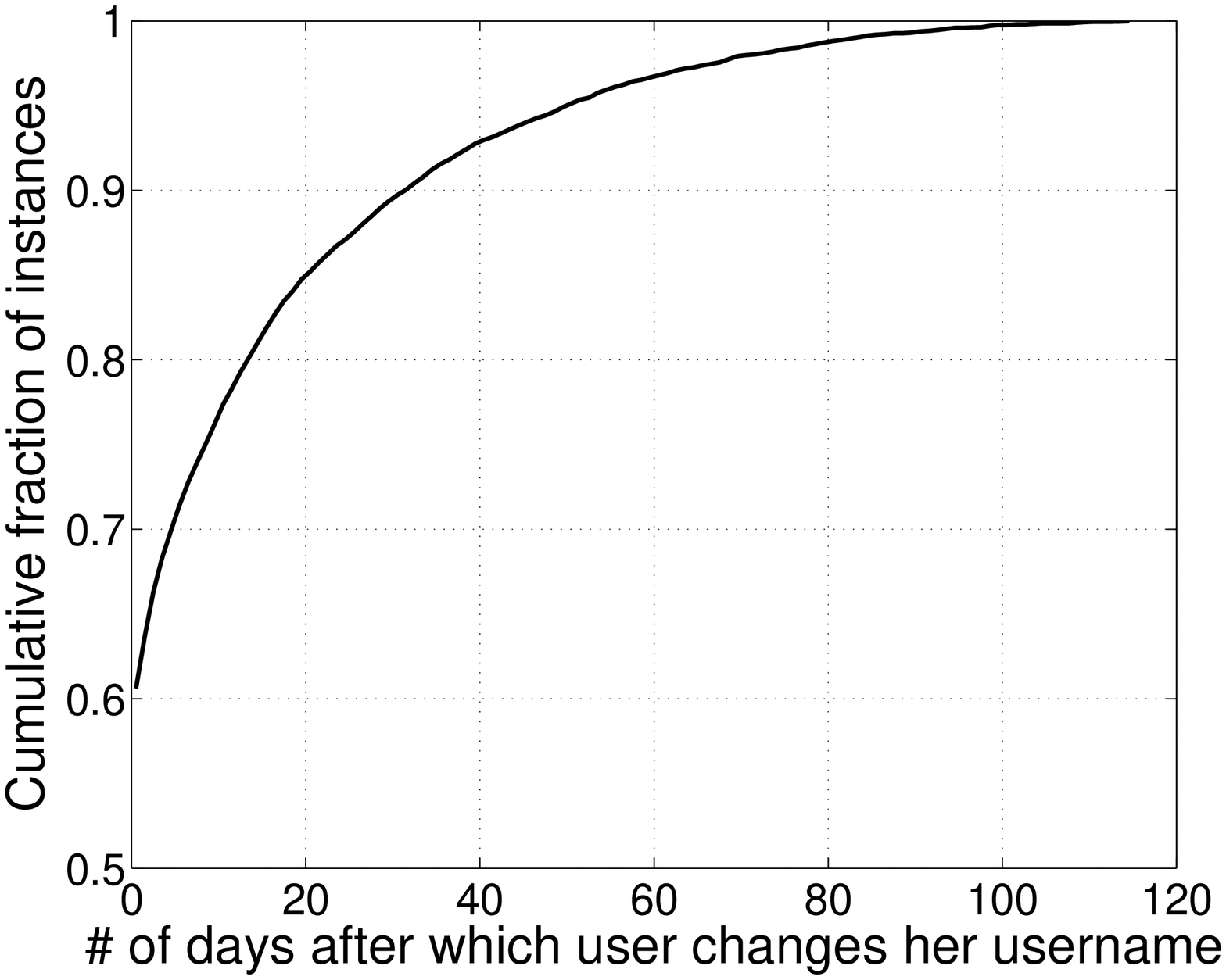}\label{fig:freq}
   }
   \quad
      \subfigure[]{
   \includegraphics[scale=0.23]{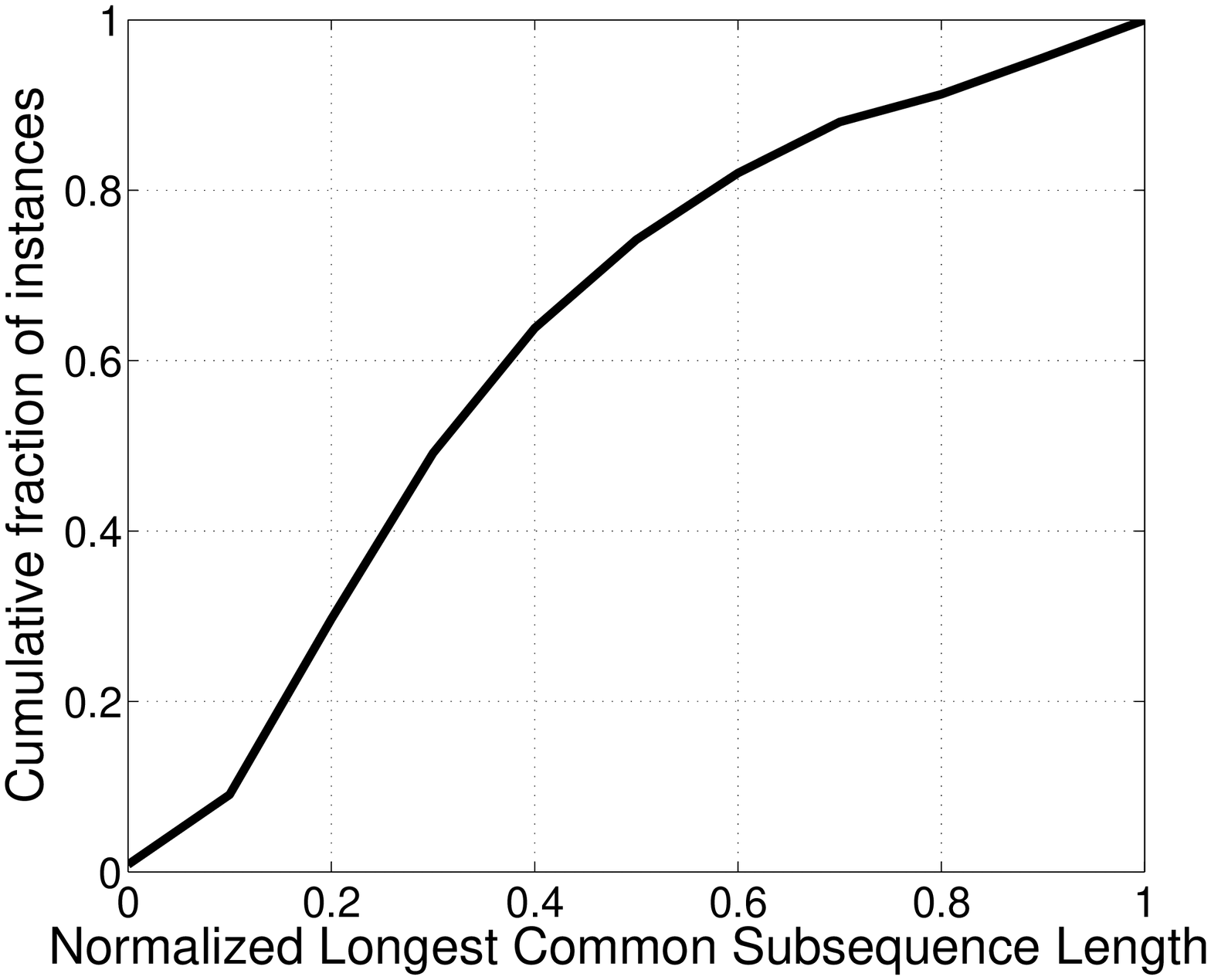}\label{fig:distance1}
   }

   \caption{(a) Username change frequency distribution (b) Cumulative distribution of username change instances vs \# of days after which users change usernames (c) Cumulative distribution of username change instances vs the longest common subsequence length between old and new usernames. We observe that few users change usernames frequently and most username changes (61\%) happen within 24 hours of earlier change (0 day). For 75\% of username change instances, new usernames are less similar to old ones.}
      \label{fig:how}

\end{figure*}
\section{Username changing behavior} \label{In-Depth}
We now explore the characteristics of username changing behavior on Twitter. We study frequency and patterns of username change and consequences to users' old usernames. 
\subsection{Frequency of username change}
We intend to understand if users change their usernames once or repeatedly. Out of 10K users, fifteen-minute scan records 2,698 users who change their usernames at least once after Nov 22, 2014. We plot distribution of users v/s number of times they change their usernames (see Figure~\ref{fig:frequency1}). Most users (1,474) change their usernames once during our observation period (Nov 22 - Mar 19), however few users change about 30 times. One user changes her username 96 times in 117 days. On manual inspection, the user~\footnote{https://twitter.com/intent/user?user\_id=796101102} seems to have malicious intentions with half completed tweets, tweets with same text, and frequent posts in short duration. We further investigate the time after which users change their usernames. We plot cumulative distribution of username change events with the number of days after which users trigger a username change event (see Figure~\ref{fig:freq}). We observe that around 61\% of username change events are triggered within 24 hours (zero day) of the previous username change. We conclude that a few users frequently change usernames within short period of time. \\
\indent We now inquire what usernames users switch to. Do they change to random usernames or to usernames similar to their dropped username? We calculate longest common subsequence length between consecutive usernames picked by the user in a username sequence. For instance, if a user $i$ chooses username $u_{i1}$ at time $t_{1}$, username $u_{i2}$ at time $t_{2}$ and username $u_{i3}$ at time $t_{3}$ (where $t_{1}$ $\leq$ $t_{2}$ $\leq$ $t_{3}$), then longest common subsequence lengths are calculated for \{$u_{i1}$, $u_{i2}$\} and \{$u_{i2}$, $u_{i3}$\}. We then normalize the length scores for each username pair with the length of the earlier username used by the user. We choose longest common subsequence matching in comparison to edit distance because the prior metric captures maximal alignments of characters between two usernames while later counts misalignments. We intend to penalize less for addition / re-arrangement of characters during a new username selection. Figure~\ref{fig:distance1} shows the cumulative distribution of username change events versus the longest subsequence matching length between the usernames (old and new) associated with the event. Around 75\% of username change instances demonstrate that the new username is less similar to the old username (length $\leq$ 0.5) and around 10\% instances have new usernames highly similar and derived from the old usernames (length $\geq$ 0.8). We therefore conclude that most users tend to pick a username which is less similar to the earlier username.
\vspace{-3mm}
\subsection{Patterns of username selection} \label{patterns}
Researchers have observed that most users choose same or similar usernames across multiple OSNs, owing to the human memory limitation to remember username for each OSN, they register to~\cite{ZafaraniL13,perito}. However, on Twitter, they need not remember their old usernames and therefore, username selection might be completely random and un-related to their earlier username. We observe the same in earlier section where 75\% of username change instances have less similar usernames. With this observation, we speculate that when a user chooses a username, she favors a username different from the earlier one.  \\
\indent We analyze 2,698 users of fifteen-minute dataset and found that out of 1,224 users who change username twice or more, 779 users (64\%) choose new usernames whenever they change. However, 445 users (36\%) choose to pick at least one of their dropped usernames again. For instance, a user $i$ picks a username $u_{i1}$ at time $t_{1}$, change to username $u_{i2}$ at $t_{2}$ and then switches back to username $u_{i1}$ at time $t_{3}$. We term such instances as \emph{username switching-back instances / events}. For such instances, $u_{i1}$, the username to which user $i$ switches back to, is termed as \emph{revisited-username} and is denoted by $u_{ir}$. Username(s) picked by the user in the meantime, collectively are termed as \emph{transit usernames}. Username $u_{i2}$ is an instance of transit username. We now extract patterns in which users choose to pick a revisited-username via the following methodology. \\
\indent We analyze a user's username sequence, for instance, \{$u_{i1}$, $u_{i2}$, $u_{i3}$, $u_{i2}$, $u_{i4}$, $u_{i3}$\}. Such a username sequence can be re-written as \{$u_{i1}$, $u_{ir1}$, $u_{ir2}$, $u_{ir1}$, $u_{i4}$, $u_{ir2}$\}. The user chooses to revisit two of her past usernames ($u_{i2}$, $u_{i3}$) denoted by ($u_{ir1}$, $u_{ir2}$). We extract three patterns from the given sequence -- \{$u_{ir1}$, $u_{ir2}$, $u_{ir1}$\}, \{$u_{ir2}$, $u_{ir1}$, $u_{i4}$, $u_{ir2}$\}, and  \{$u_{i1}$, $u_{ir1}$, $u_{ir2}$, $u_{ir1}$\}. Pattern \{$u_{ir1}$, $u_{ir2}$, $u_{ir1}$\} represents a pattern of length 3 where user chooses an immediate dropped username. Pattern  \{$u_{i1}$, $u_{ir1}$, $u_{ir2}$, $u_{ir1}$\} represents a pattern of length 4 where user chooses an immediate dropped username and contains a sub-pattern  \{$u_{ir1}$, $u_{ir2}$, $u_{ir1}$\}. Pattern \{$u_{ir2}$, $u_{ir1}$, $u_{i4}$, $u_{ir2}$\} of length 4 which represents that user chooses to revisit a username used by her, two username change events earlier. We extract all patterns and sub-patterns of different lengths where user chooses to reuse her past username, from 445 user sequences (see Table~\ref{fig:pattern}). \\
\indent  We observe that (sub)pattern \{$u_{irj}$, $u_{in}$, $u_{irj}$\} is most  frequent among username switching-back instances than others. Users favor most-immediate dropped username when they wish to choose from their past usernames. We think that users hop to most recently dropped username again, either because their other past usernames are taken, they may not remember their rest past usernames or they remember the benefits of their most recently dropped username. We believe that such a characteristic can help Twitter to design customized username recommendation feature. Twitter can recommend its users the most beneficial username among their past usernames based on followers count gain (Section~\ref{gainloss}), activity and other metrics, rather than they choosing recently dropped usernames again, owing to lack of information about suitability of their other past usernames.
\begin{table}[htbp]
\vspace{-3mm}
\begin{center}
\small
\begin{tabular}{|p{4cm}|p{2.5cm}|}
\hline
\raggedright{{\bf Pattern / Sub-pattern}} & {\bf{\# of instances }} \\ \hline
$u_{ir1}$ - $u_{i2}$ - $u_{ir1}$ &  228  \\ \hline
$u_{i1}$ - $u_{ir1}$ - $u_{i2}$ - $u_{ir1}$ & 104 \\ \hline
$u_{ir1}$  - $u_{i2}$  - $u_{i3}$  - $u_{ir1}$  & 54  \\ \hline
$u_{ir1}$  - $u_{ir2}$  - $u_{ir1}$  - $u_{ir2}$  & 18 \\ \hline
$u_{ir1}$  - $u_{i2}$  - $u_{i3}$  - $u_{i4}$  - $u_{ir1}$  & 12\\ \hline
$u_{i1}$  - $u_{ir1}$  - $u_{i2}$  - $u_{i3}$  - $u_{ir1}$  & 8 \\ \hline
$u_{ir1}$  - $u_{i2}$  - $u_{ir1}$  - $u_{i3}$   - $u_{ir1}$  & 9 \\ \hline
$u_{ir1}$  - $u_{ir2}$  - $u_{i3}$  - $u_{ir2}$  - $u_{ir1}$  & 3\\ \hline
$u_{i1}$  - $u_{ir1}$  - $u_{ir2}$  - $u_{ir1}$  - $u_{ir2}$  & 9 \\ \hline
$u_{ir1}$  - $u_{ir2}$  - $u_{i3}$  - $u_{ir1}$  - $u_{ir2}$  & 2 \\ \hline
\end{tabular}
\caption{Popular patterns of revisited-username selection. Most users switch back to recently dropped username.}
\label{fig:pattern}
\end{center}
\end{table}
\vspace{-6mm}
\subsection{Squatted usernames}
Till now, we explore the properties of new usernames picked by users. We are also interested in examining the properties of old dropped usernames i.e. the usernames which users vacate to pick new usernames. On Twitter, there are four pools to which an old username  can belong to -- \emph{free-username pool}, \emph{taken-username pool},~\footnote{https://support.twitter.com/groups/51-me/topics/205-account-settings/articles/14609-changing-your-username} \emph{suspended / deactivated-username pool}~\footnote{https://support.twitter.com/articles/15348-my-account-information-is-already-taken\#deactivatedaccount} and \emph{squatted-username pool.}~\footnote{https://support.twitter.com/articles/18370-username-squatting-policy} A username $u_{io}$ belongs to a free-username pool if no one else uses it on Twitter. If another user selects username $u_{io}$ as her own, the username moves to taken-username pool. If the user with username $u_{io}$ deactivates her account or is suspended by Twitter, the username is blocked forever (for now) and is not available to anyone, which thereby moves the username in suspended / deactivated-username pool. If an inactive user profile registers her account using $u_{io}$, in order to block or preserve that username, and not to allow others to use it, the username is said to belong to squatted-username pool.  \\
\indent Squatted usernames on OSNs have been investigated as a challenge in literature by law researchers~\cite{ramsey2010brandjacking,curtin2010name}. Researchers analyzed trademark infringement cases on social media like Twitter. We were curious to know if cybersquatting exists on Twitter in the form of username squatting. We could not consider cases of trademark infringements, because of the lack of ground truth, where companies filed a trademark infringement report to Twitter. Therefore, we check for generic users, if users' past / dropped usernames are reserved by inactive Twitter user profiles. \\
\indent For our fifteen-minute dataset, we observe that for around 7\% of 2,698 users, at least one of their vacated usernames are blocked by inactive Twitter profiles (either created by themselves or others) who either show no activity (i.e. no tweets) or have zero followers. We think that inactive profiles may have been created to avoid slip of past usernames in Twitter's free-username pool. Twitter considers such blocked usernames as `squatted' usernames. However, according to Twitter username squatting policy, squatted or inactive usernames are not released, unless the username causes trademark infringement. In such scenarios, common Twitter users are suggested to choose a modified version of their wished username. We suggest that Twitter should implement a username request feature, where a user can put her request for a username, if the username is not available and as soon as the username is vacated, the user can be notified. This way, Twitter can avoid squatted usernames and users can use their requested username. Further, investigating if a user is blocking her own past usernames via inactive profiles, can help Twitter understand her malicious intentions.
\section{User characteristics} \label{user}
We now attempt to understand the characteristics of users who opt to change their usernames. Do users with high popularity or activity change their usernames frequently? Do new users change their usernames more frequently than old users? Do they differ from users who never change their usernames? We answer each of these questions now.
\subsection{In-Degree centrality}
On Twitter, users tweet, reply or indulge in conversations with their username. Changing usernames by a popular user may lead to confusion among her followers or may lead to loss of tweets in case someone else picks the username. In such a scenario, we speculate that users with high centrality / in-degree would like to avoid any username change. We use fifteen-minute dataset of 2,698 users and measure the in-degree of the users (see Figure~\ref{centrality}). We correlate in-degree of a user with the frequency of username changes. We remove an outlier user who changes her username 96 times (in-degree - 14.3K). We observe that number of times a user changes her username is weakly and positively correlated to the in-degree of the user (Pearson correlation: 0.0083). We therefore conclude that irrespective of the popularity, a user may still wish to change the username multiple number of times. We suspect that popular users change interests, and put opinions about recent events frequently and therefore, to reflect the same, change their username multiple times.

\subsection{Activity}
\begin{figure}[htbp] 
   \centering
   \subfigure[]{
   \includegraphics[width=0.5\linewidth]{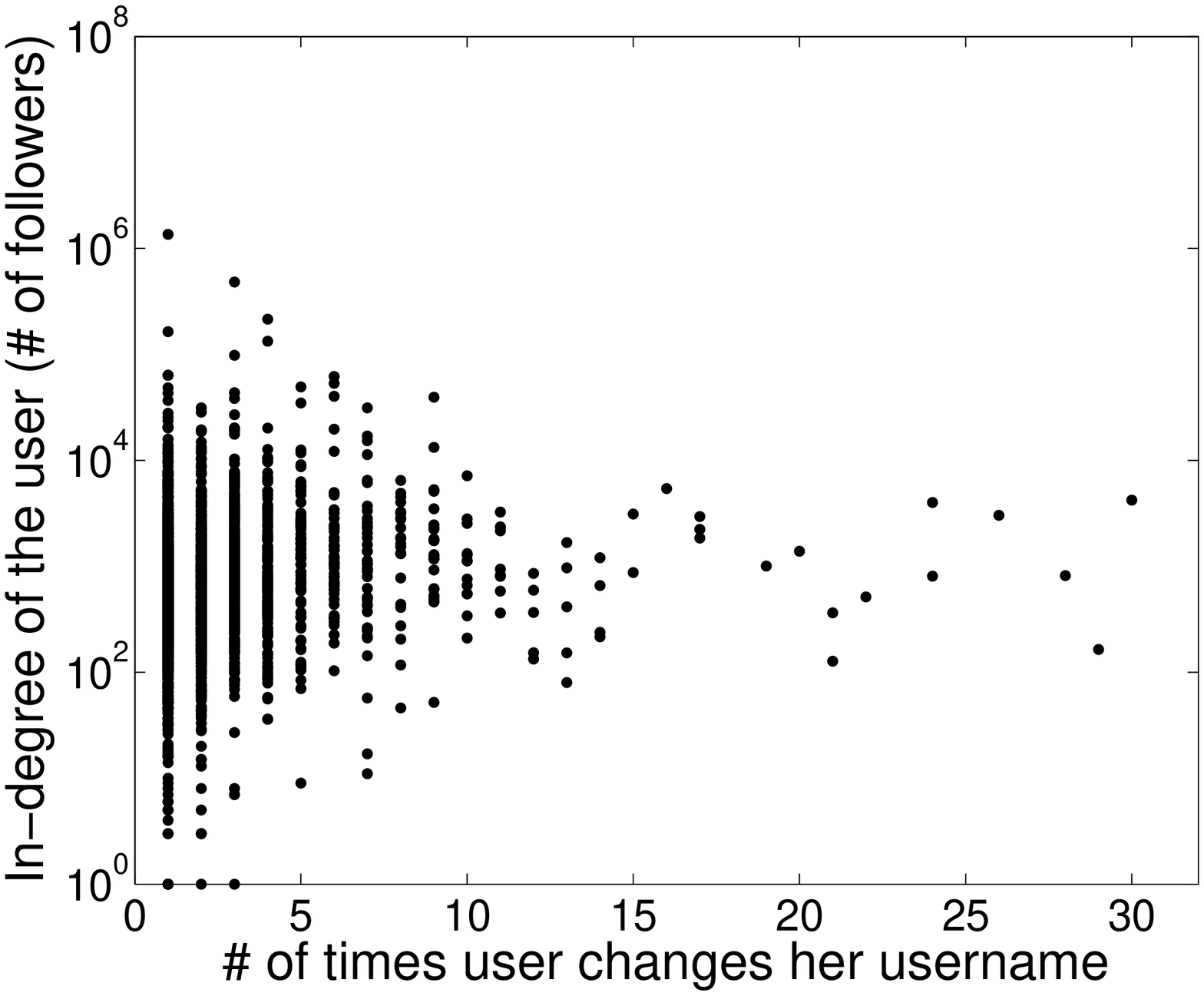} \label{centrality}
   } 
    \subfigure[]{
    \includegraphics[width=0.5\linewidth]{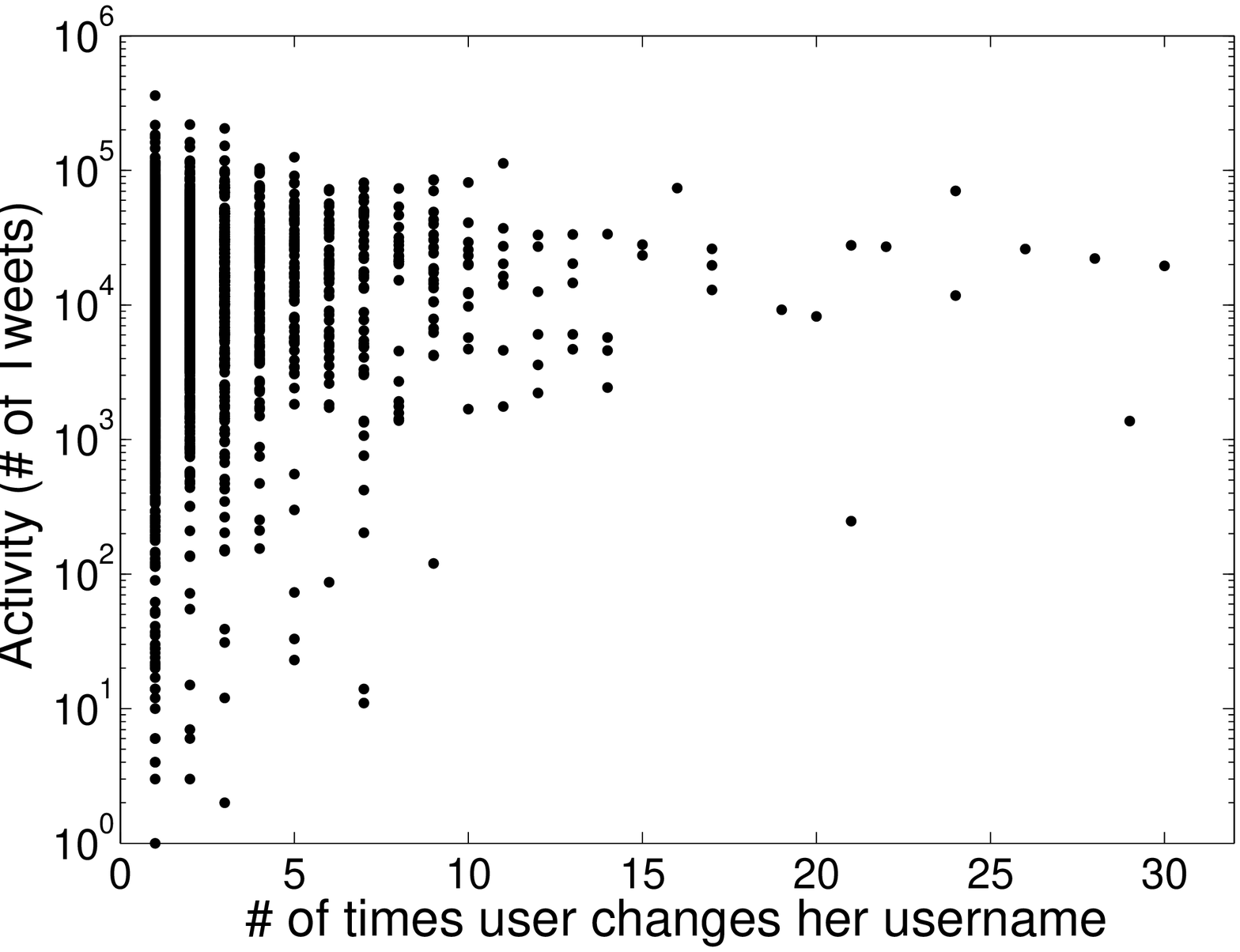} \label{statuses}}
   \caption{ (a) user's in-degree distribution, (b) user activity distribution versus frequency of username changes. We find that in-degree centrality and activity of a user are weakly correlated with the number of times she changes her username.}
   \label{fig:multiple}
\end{figure}

An active user on Twitter, who engages herself in conversations and group chats, may change her username less frequently to avoid confusion during tagging / replying in a tweet. We therefore speculate that active users change their usernames less frequently. We analyze 2,698 users and measure their activity with the number of tweets they create. We remove an outlier user who changes her username 96 times (tweet count - 8,741). Figure~\ref{statuses} shows the tweet distribution of users with the number of times they change their usernames. We observe a weak and positive correlation between the two (Pearson correlation: 0.0322). We infer that users who actively post on Twitter change their usernames frequently for possible reasons such as to gain traction or change behavior during trending topics (explored in Section~\ref{reasons}).

\subsection{Age of the account}
\begin{figure}[htbp] 
   \centering
   \includegraphics[width=0.9\linewidth]{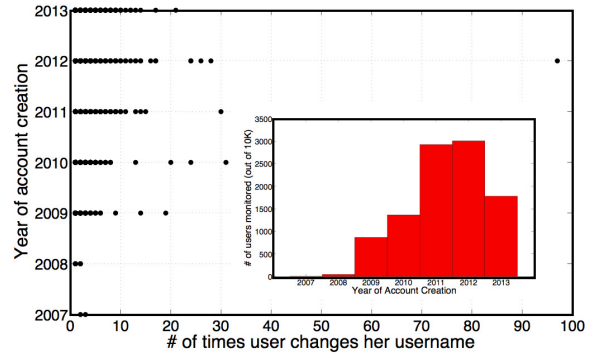} 
   \caption{shows age of the user profile v/s number of username changes and distribution of 10K users monitored. We observe a weak correlation between the account age with the frequency of username change. }
   \label{age}
\end{figure}

Users who registered themselves on Twitter, long time ago might have chosen most stable and beneficial username for themselves than users who have registered recently and are still in exploratory stage. We examine if old user accounts engage themselves in username changing behavior or only new users change their usernames multiple times. Figure~\ref{age} shows the age of the account distribution for 2,698 users, with the number of times users change their usernames. Inset graph shows distribution of monitored 10K users across year of their account creation. We observe negative and weak correlation between the age of the Twitter account and the frequency with which the account changes username (Pearson correlation: -0.0798). Negative correlation implies that older accounts change their usernames less frequently, however, is not necessarily true for all old accounts. A 2009 account changes her username 19 times, while a 2010 account changes her username 31 times. We therefore infer that irrespective of the age of their account on Twitter, users change their usernames.

\subsection{Normal v/s Changing users}
\begin{figure*}[htbp] 
   \centering
   \subfigure[In-degree distribution of users]{
   \includegraphics[scale=0.24]{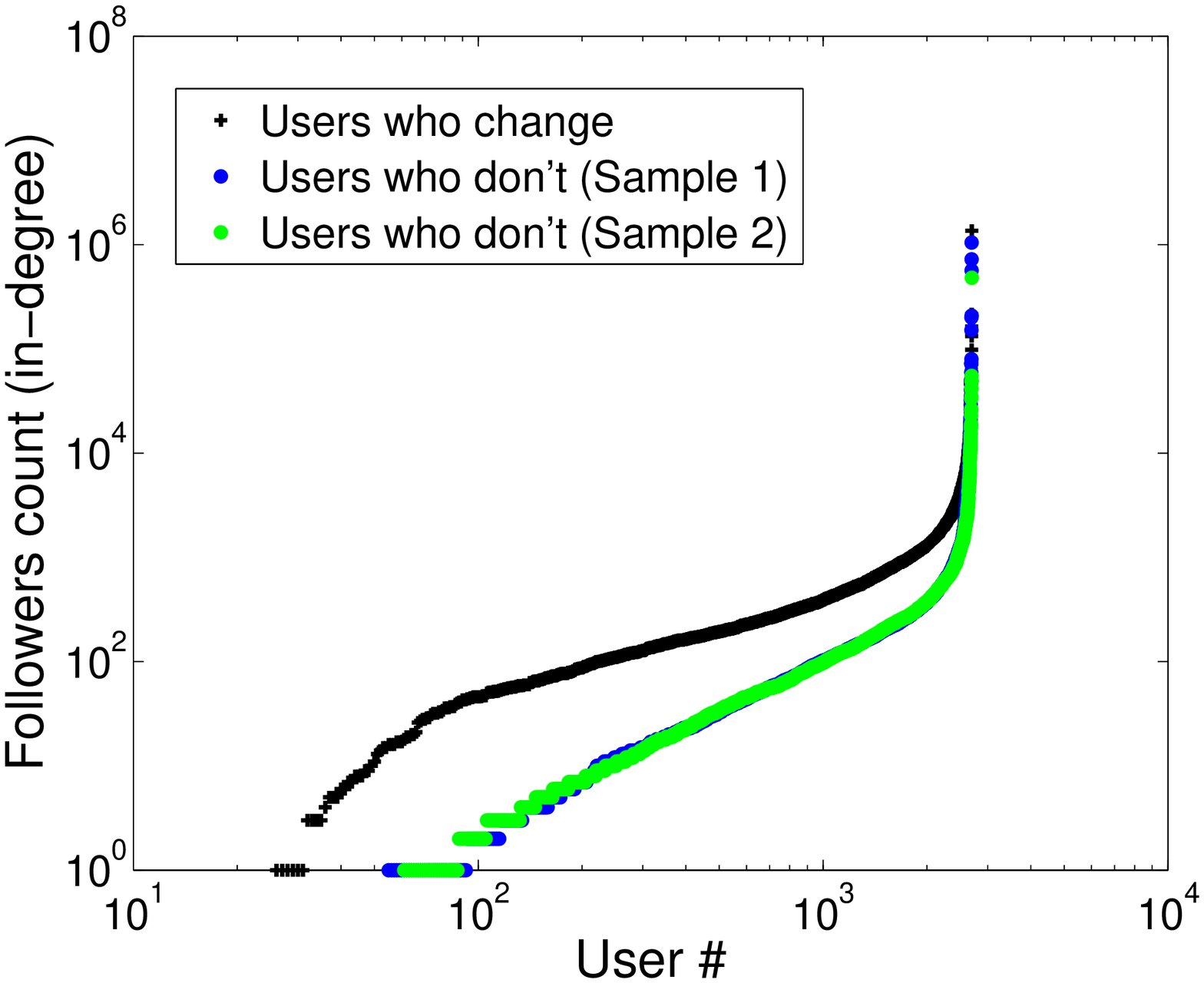} \label{in-deg}
   }
   \subfigure[Out-degree distribution of users]{
    \includegraphics[scale=0.24]{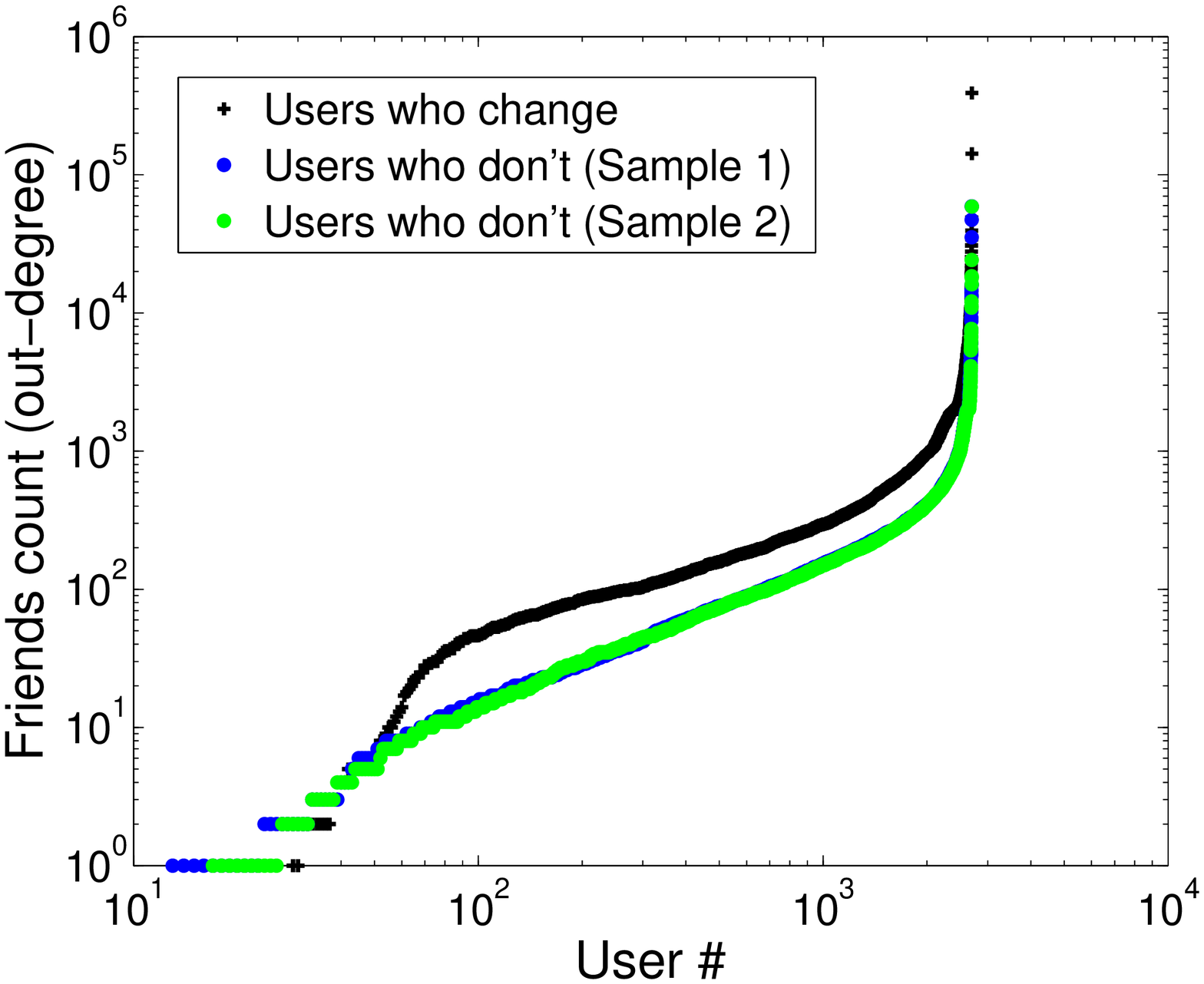} \label{out-deg}}
   \subfigure[Activity distribution of users]{
    \includegraphics[scale=0.24]{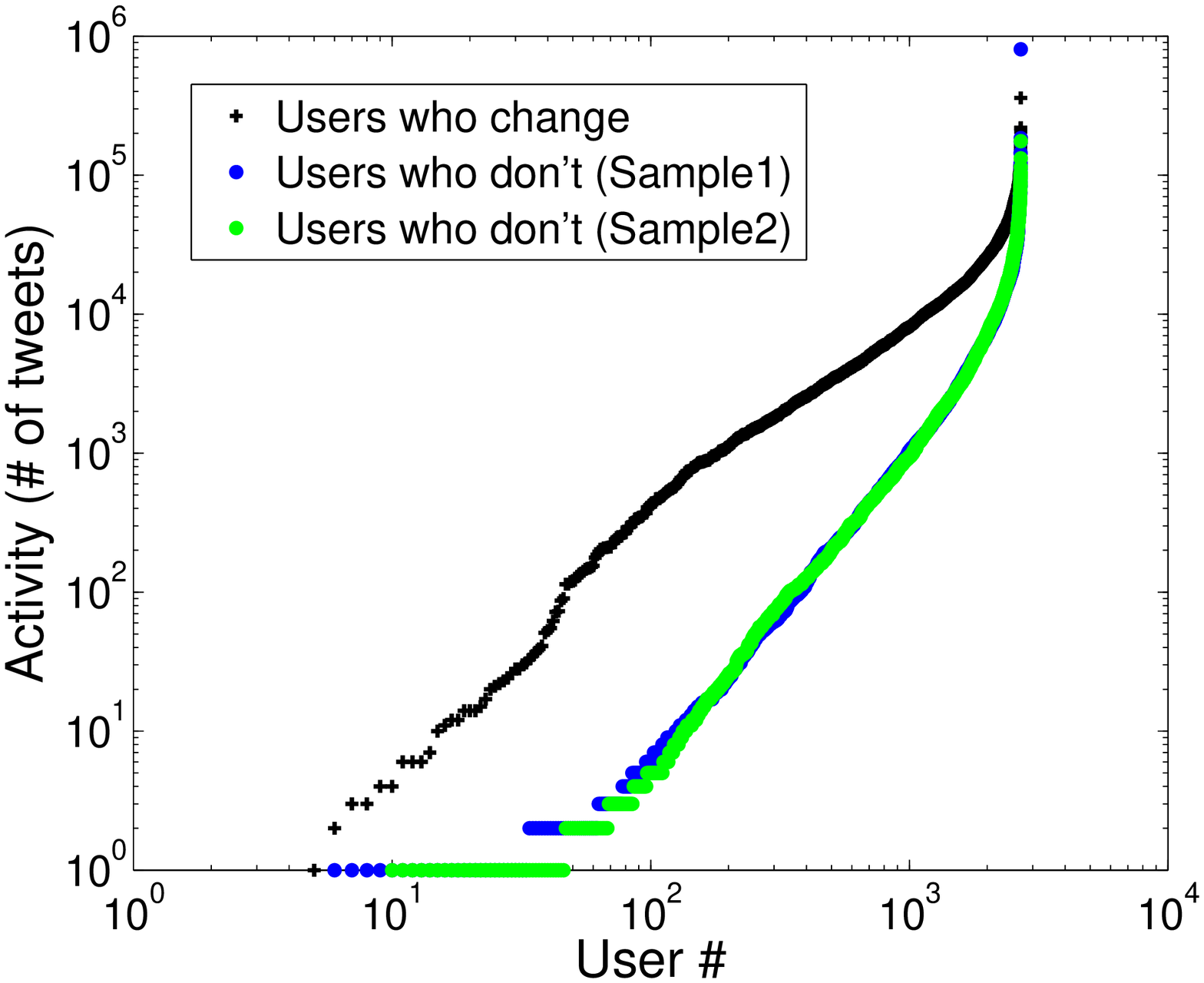} \label{status}}
    
   \caption{shows a comparison between 2,698 users who changed their usernames with two random samples of 2,698 users who never changed their usernames. Users who change usernames demonstrate superiority in terms of popularity, and activity and users being followed as compared to users who don't. }
   \label{comparison}
\end{figure*}

As we observe for 8.7M users, 10\% of users change their usernames over time. A large proportion (90\%) still does not engage in this behavior. We now explore the similarities and differences between the properties of users who do not change their usernames with users who do. We compare in-degree, out-degree and activity of 2,698 users with characteristics of two random samples of size 2,698 users,~\footnote{We take two random samples to justify the generalizability of observations. } extracted from 8.7M users who have not changed their usernames during Oct 1 - Nov 26 (see Figure~\ref{comparison}). We observe that users who change their usernames demonstrate higher popularity and activity and actively follow other Twitter users as compared to users who do not bother to change their username. Note that, both random samples do not differ in their properties but differ from the users who change their usernames over time.

\section{Plausible reasons} \label{reasons}
We now examine the reasons why users opt for a username change by analyzing fifteen-minute dataset. We validate the reasons with the interactions we have with the users we monitor, as discussed in this section.
\subsection{Space gain}
\begin{figure}[htbp] 
   \centering
   \includegraphics[scale=0.23]{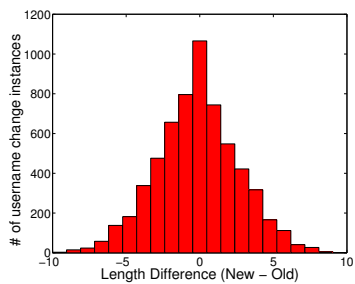} 
   \caption{Length difference between new and old username. We observe 44\% username change instances where new username is shorter than the old one. Space gain can be one of the reasons (but not only) for changing username. }
   \label{fig:length}
\end{figure}

Twitter allows its users to post 140 characters in a tweet. Long usernames might allow less space to convey content, while short usernames might give the liberty to add more content, hashtags or URLs~\cite{chhabra:phi.sh/ocial:-the-phishin:2011:yuqfj}. Therefore, users with long old usernames may change to short new usernames to benefit from space gain. We calculate the length difference between consecutive usernames in a username sequence i.e. between old and new username of a user and plot the distribution (see Figure~\ref{fig:length}). We observe that out of 6,132 instances of username change by 2,698 users, 2,687 instances exist  where the new username is shorter than old username (44\%), 2,378 instances exist where new username is longer than the old username (39\%), 1,067 instances witness no length change between old and new username (17\%). Therefore, around half the population changes usernames to benefit from space gain, however the other half choose same or longer length username. A female responder from the user survey (discussed in Section~\ref{Discussion}) mentions that she changes her username to gain space. We conclude that space gain is a possible (but not only) reason for users to change their usernames.

\subsection{Gain followers or avoid loss of followers}\label{gainloss}
As observed in Section~\ref{patterns}, 445 users choose to switch back to any of their earlier usernames. We examine the possible reasons for users to change and switch back to any of their past usernames (revisited-username). In real-world, researchers found that users keep the identity which causes maximum benefits, in terms of friends and reputation~\cite{mcfarland2005motives}. We validate the finding in our online scenario. We examine fifteen-minute dataset and observe three plausible reasons of why users switch to an earlier used username -- loss due to transit usernames, gain due to revisited-username, or both. We discuss each of the reasons now. 
\vspace{-2mm}
\subsubsection{Loss or no gain via transit-usernames}
\begin{figure}[htbp] 
   \centering
   \includegraphics[scale=0.38]{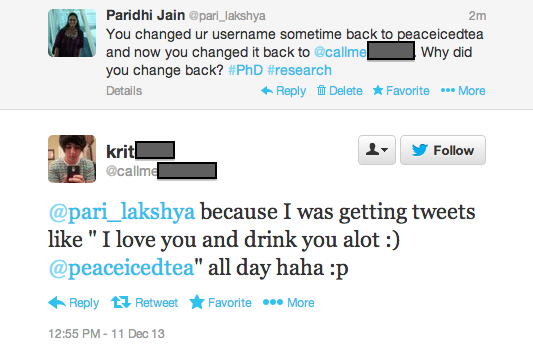} 
   \caption{User changes username to pick a brand's old username (transit-username), receives no followers gain, and irrelevant tweets, and hence, switches back to his earlier username.}
   \label{fig:spam}
\end{figure}

We quantify loss in terms of loss of followers for the user. We explore if loss or no gain of followers due to the transit usernames, collectively, prompts users to switch back to the revisited-username. We calculate the follower count difference between the time user first uses the revisited username to the next time the user uses the revisited username. We observe that for 158 users (36\%), transit-usernames collectively cause a loss or no gain of followers (follower count change $\leq$0). For instance, a user ID 76xx33242 with 3,288 followers, might have experienced loss of 674 followers due to her transit-username, and therefore might have switched back to her earlier username after 11 days. Maximum observed loss of followers due to transit usernames is -6,118 for a user with 215,084 followers earlier. Further, 26\% of 158 users switch back to their earlier username within 24 hours. We also talk to some of the users by tweeting them to know why they switch back to earlier username. A user responds that he gains no followers and receives a lot of irrelevant tweets with the transit-username and therefore he changes back to the earlier username (see Figure~\ref{fig:spam}).

\vspace{-2mm}
\subsubsection{Gain via revisited-username} 
We quantify gain in terms of the gain of followers for the user. We explore whether the user wishes for revisited-username again because it has caused her a gain in follower count in the past, and therefore has helped her to increase her popularity. We calculate the follower count difference between the time user picks the revisited username to the time when she drops it and initiate another username change event. We observe that for 211 users (47\%), revisited-username has gained followers. For instance, user ID 56xx15159 gains 51 more followers by choosing a revisited-username for the first time, gains 1 follower due to transit username and gains 63 followers when she switches back to the revisited-username. We, therefore, infer that the experience of the user with her past usernames governs her decision on selection of revisited-username. \vspace{-2mm}
\subsubsection{Loss via transit \& gain via revisited username}
A mix of both reasons explained earlier is also observed in our dataset. 41 users (9\%) experience both loss of followers due to transit-usernames and gain of followers due to the revisited-username. A male responder from the user survey (discussed in Section~\ref{Discussion}) mentions that he gains followers due to revisited-username and lose followers due to transit usernames, hence switches back. 

\indent We therefore infer that, most users do not stake their popularity for a new username and will change immediately if they observe a fall in their followers. To choose a revisited-username among the past usernames, users prefer most recently dropped username as we observed earlier (see Table~\ref{fig:pattern}). We reason this behavior with the limited memory and gauging capability of users to understand which usernames in their past caused them maximum benefit, further with no such help from Twitter. Twitter does not provide any support to help users to measure gain or loss of followers as they change usernames over time. We suggest that Twitter can develop a better understanding on which usernames a user should choose, based on the associated benefits and losses observed in the past, thereby can suggest better and tailored usernames to its users.\\
\indent On the other side, there are instances where users have gained followers due to transit-usernames or have lost followers due to revisited-username. For 287 users (64\%), transit usernames cause a gain of followers ($\geq$ 1). A user with 1 follower earlier, gains 106,970 followers within 3 days with her transit username, and yet changes back to her earlier username.Further, only 6\% of such users who gain via transit-username, switch back to the earlier username within 24 hours. We think that users who gain via transit usernames intend to wait till they achieve maximum benefit of followers and then switch back to earlier username due to other reasons explained in the section. 

\begin{table*}[htbp]
\begin{center}
\small
\begin{tabular}{|l|l|l|l|l|p{3cm}|}
\hline
\raggedright{{\bf ID}} & {Scan - I} & {Scan - II } & {Scan - III}  & {Group} & {Date of observation when user holds the tracked username} \\ \hline
12xx62463x & \textbf{CollaGe\_InFo} & Dictionary\_ID & Dictionary\_ID & Sajan Group & 2013-04-01  \\ \hline
11xx79686x & DaiLy\_GK & \textbf{CollaGe\_InFo} & Geo\_Account & Sajan Group & 2013-10-02   \\ \hline
95xx1822x & Geonewspak9 & DictioNary\_GK & \textbf{CollaGe\_InFo} & Sajan Group & 2013-10-25 \\ \hline
19xx56472x & - & - & \textbf{CollaGe\_InFo} & Sajan Group & 2013-12-04 \\ 
\hline
\hline

60xx2762x & \textbf{Peshawar\_sMs} & MoBile\_TricKes  & BBC\_PAK\_NEWS & Sajan Group & 2013-04-08 \\ \hline
11xx37099x & Vip\_Wife & \textbf{Peshawar\_sMs}  & UBL\_Cricket & Sajan Group & 2013-10-25  \\ \hline
28xx1645x & NFS002cric & NaKaaM\_LiFe& \textbf{Peshawar\_sMs} & Sajan Group & 2013-11-08 \\ \hline
\hline
70xx9502x & \textbf{FuNNy\_SardaR} & MaST\_DuLHaN & KaiNaT\_LipS & Khan Group & 2013-04-01 \\ \hline
99xx9356x & SaIrA\_JoX &  \textbf{FuNNy\_SardaR} & MaST\_DuLHaN & Khan Group & 2013-10-02  \\ \hline
12xx73970x & - & - &  \textbf{FuNNy\_SardaR} & Khan Group &   2013-12-04 \\ \hline
\end{tabular}
\caption{Rotational use of a username among the members of a group in Twitter. We observe that different users pick the same username at different times, which might intend towards promotion of the username.}
\label{tab:rotation}
\end{center}
   \vspace{-3mm}

\end{table*}

\subsection{Rotational use of a username in a group} \label{rotation}
Owing to limited number of users in fifteen minute dataset who changed their usernames, we use our four scans and seed set of 8.7M users for this analysis. We observe that a few user profiles, who change their usernames, belong to or follow a group and their past usernames are picked by other user profiles following / belonging to the same group. Table~\ref{tab:rotation} shows two such groups and the rotation of a username among the profiles, as observed in four scans. Username `Collage\_InFo' is used by a user with user id 12xx6246xx as recorded in Scan-I. The same username is later used by user with user id 11xx7968xx as observed in Scan-II, followed by user 95xx182xx in Scan-III and by different user later. All users who use a particular username at different timestamps, claim that they belong to a group named as `Sajan Group', either in their name attribute or in their bio attribute. Similar behavior is observed with username `Peshawar\_sMs' and `FuNNy\_SardaR'. We observe 70 such usernames in four scans and seed set, which are picked by different users at different timestamps. We speculate that either user profiles who keep the same username belong to the same real-world user, or the intention is to popularize the username itself. We think that users change usernames in order to let other users in the group use the username.~\footnote{We do not tweet such users to avoid alarming them in case they have malicious intentions.} 
\subsection{Other possible reasons}
\begin{figure}[htbp]
\centering
   \subfigure[]{
   \includegraphics[scale=0.35]{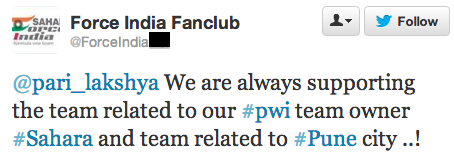}\label{fig:changes1}
   }
   \quad
  \subfigure[]{
   \includegraphics[scale=0.35]{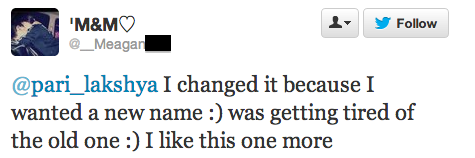}\label{fig:changes2}
   }
   \caption{Username change (a) due to change of events over time, (b) due to boredom.  }
      \label{fig:how}
\end{figure}

Other reasons we either inspect manually from our dataset or learn by tweeting and asking the users we monitor are --
\begin{itemize}
\item
\vspace{-3mm}
\textbf{Change of username identifiability} -- Few users in our dataset change usernames to reverse the identifiability of the usernames i.e. either to make them personal or anonymous. For instance, a user named `loried ligarreto' changes her username from `loriedligarreto' to `sienteteotravez' (feel again in English) implying that user intends to make her username anonymous. In other instances, we observe users who previously pick less identifiable usernames, make them personal later. For example, a user named `rodrigo' changes her username from `unosojosverdes' (green eyes in English) to `rodrigothomas\_', thereby implicating that user wishes to associate her real identity to her username.
\item
\textbf{Change of Events} -- A user responds that she represents Sahara India FabClub. She has supported Sahara's Pune Warriors team in IPL event with username `pwifanclub' and then Sahara F1 team with username `ForceIndia@!' and therefore has changed her username (see Figure~\ref{fig:changes1}).
\item
\textbf{No specific reason} -- Few users respond that they change their usernames without any specific reasons, that they get bored of the earlier one (see Figure~\ref{fig:changes2}).
\end{itemize}

\section{Applications} \label{Applications}
We now discuss the application domains where recorded past usernames can be helpful. We show that past usernames of a user can help Twitter to come up with customized username suggestions and can help to create a unified social footprint of a user by locating her usernames on other OSNs. Tracking a username and user profiles who choose a particular username can help in correlating user profiles to a single real-world person. 


\subsection{Username suggestion}
We now know that users change usernames frequently on Twitter. They change to non similar usernames as compared to earlier ones, which help them to increase their popularity, gain more space, maintain their anonymity (identity), or avoid boredom. We suggest that the understanding of such facts can help Twitter to suggest usernames to its users, they may wish to change to, by recording and analyzing their past usernames and assessing their suitability to the users. Suggestions can further be customized based on users' username selection patterns. We think that suggesting usernames to users may improve users' experience with Twitter, in terms of better visibility, newer connections and higher relevant interactions with newly suggested usernames. 

\subsection{Identity resolution}
With a user registered on multiple OSNs with varying characteristics, it is difficult to find her identity on multiple OSNs. The problem of finding and resolving a user's multiple identities across different OSNs is termed as ``Identity resolution in OSNs" \cite{jain2013seek}. Till now, researchers have suggested to use most recent profile attributes of users to help in identity resolution~\cite{ZafaraniL13, perito,Irani,Malhotra, goga2013exploiting}, however little research explores the potential of past profile attribute changes, specifically usernames, in solving identity resolution in OSNs.\\
\indent We hypothesize that a user might use any of her past usernames from Twitter as her username on other OSNs, to remember selected chosen usernames for all OSNs. We test this hypothesis with users who change their username at least once as recorded in fifteen-minute dataset and the four scans. To test the hypothesis, we need users' other OSNs usernames, to compare them with users' past usernames. However, we don't have direct access to such information. Therefore, we use self-identification method~\cite{jain2013seek} to search and find users' usernames on other OSNs. We extract their self-claimed account (username) on other OSNs, via their URL attribute value. For instance, a user mentions `$http://www.facebook.com/xyzabc$' as her URL attribute thereby exposes and self-identifies her Facebook username as `$xyzabc$'. With this methodology, we find 214,636 out of 853,827 users and 993 out of 2,698 users, mention their other OSN accounts via URL attribute on Twitter. We then compare extracted usernames on other OSNs with  past usernames of the user on Twitter.~\footnote{Few users change URL temporally to direct to their other OSN accounts or new account on an earlier referred OSN.}\\ \indent We find 9\% of 214,636 users and 8\% of 993 users choose at least one of their past usernames to register on other OSNs. Therefore, an exact string search with user's past usernames can help locating her other OSNs accounts. Note that, we might have missed users for whom the hypothesis is true, because of no information on their other OSN accounts. Further, we note that a few users  change their usernames on other OSNs in the same pattern as they do on Twitter. For instance, a Twitter user changes username from ``happygul@!!" to``gulben!!!" as well as changes her Instagram~\footnote{www.instagram.com} username from ``happygul@!!" to ``gulben!!!", as observed in the URL attribute of the user. Therefore, historic usernames of a user may help to suggest her username on other OSNs and therefore, may address identity resolution in OSNs.

\subsection{Multiple identities correlation} 
\begin{figure}[htbp] 
   \centering
   \includegraphics[scale=0.37]{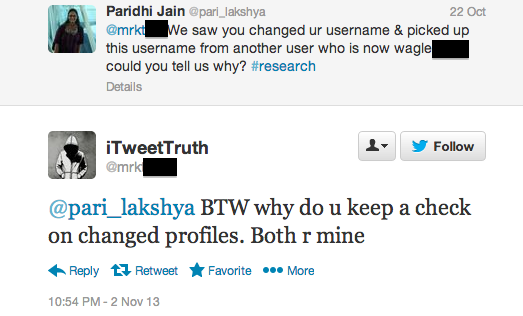} 
   \caption{User mentions that two different profiles had same username (@mrk!!!) at different times, because two profiles (@mrk!!! and @wagle!!!!) belong to the same person. }
   \label{fig:multiple_rotation}

\end{figure}

As observed in Section~\ref{rotation}, few users pick same username at different timestamps. We observe 70 such usernames in four scans and seed set, which are picked by different users at different timestamps. We inquire about few instances and send tweets to users, asking what are the possible reasons for such a behavior. Two users reply as `` Both the user profiles belong to me." (see Figure~\ref{fig:multiple_rotation}). In this way, multiple identities of a real-world user may be correlated and linked together via monitoring the user profiles and their username changes, if they use the same set of usernames over time, thus helping multiple identities correlation on Twitter. 

\section{Discussion} \label{Discussion}
In this work, we attempt to characterize, an unexplored username changing behavior on an OSN. We observe 8.7 million users at long intervals while 10,000 users at short intervals. We derive that around 10\% of users change their username owing to reasons such as space gain, gain more followers or switch with other user's username. Few users undergo a username change multiple times (96 times in 117 days) while many users change their usernames once. In some scenarios, users switch back to their past usernames. Number of times a user changes her username is weakly correlated to her popularity, activity and age of her account. We further observe that users choose their username on other OSNs out of the usernames they used on Twitter in the past. Further, monitoring which users pick the same username at different timestamps might help in understanding if different online users reflect the single real-world entity.\\
\indent We discuss quantifiable analysis, draw inferences and attempt to validate the inferences with user feedback. We wrote tweets to few Twitter users we monitored, asking them the reasons, and benefits of their username change. Only few users responded and most ignored the tweet. We then planned a survey of 10 questions customized for each user with their past usernames.~\footnote{http://nemo.iiitd.edu.in/findingnemo/user1/} We designed the survey to comprehend the reasons, benefits or losses for which users changed their usernames. We tweeted the survey to users with a message ``Changed ur username $<$\#$>$times in $\#$months? \#Help us know why. Fill $<$survey link$>$" over a week using two Twitter accounts. We tweeted the survey to only 500 users out of 2,698 users, with @tag. The reason was that our accounts got suspended multiple times because we used automated scripts to send tweets.~\footnote{We even randomized our tweeting time, but Twitter still suspended our accounts.} Three users filled the survey, two males and one female. One user responded that she changed her username to gain space, another user responded that he changed because earlier username was inappropriate and the third user changed for no specific reason. Further one male user switched back to his past username, because he gained followers due to revisited-username and lost followers due to transit usernames. Due to limited responses, we do not list other reasons for changing usernames. \\
\indent In this work, we strictly focus on the username attribute changes over time. However, we observe that apart from username, other profile attributes change over time as well. Users change their name, description, location, URL and profile picture more frequently than their username. By monitoring identity changes of a user, a user's true identity attributes, missing attributes and recency / validity of their attributes on OSNs can be estimated. We sense an immense potential in the historical data of an OSN user.
\section{Related Work} 
Researchers have examined the temporal nature of two important user attributes on an OSN namely content and network, however little has been explored about temporal changes to profile attributes of the user. Content attribute of users were studied on Twitter to understand their posting behavior. Authors suggested that at any time, users' post characteristics depends on three factors -- breaking news, posts from social friends and the user's interests~\cite{Xu:2012:MUP:2348283.2348358}. A user's interests in terms of topics she posts about, were studied to capture temporal changes in her topical interests and therefore her posting behavior~\cite{abel2011analyzing}. Depending on the user's pace of keeping unto new topics, authors built a framework to model users' temporally changing interests and used the model to build a personalized recommender system. Network attributes of a user were studied on OSNs, in terms of evolution of her friend connections~\cite{fan2011incremental,mislove2008growth}, and involvement in groups~\cite{motamedicharacterizing}. \\
\indent Profile attributes were studied temporally by \cite{5231867}, where authors crawled 2 million Myspace profiles twice over a period of a year. Authors compared the evolved honesty and accountability of the users, derived from the user's profile, content and network attributes. Authors however did not present any insights on why users preferred an identity change at the first place. Further, reasons of an identity change were explored and reasoned in the real-world by sociology and psychology researchers~\cite{mcfarland2005motives,burke2006identity}. Researchers observed 6,000 high school adolescents after a year, found identity changes of a user in terms of affiliations of the user in the group, description of herself \cite{mcfarland2005motives}. They found that network attributes of a user such as betweenness play a major role in understanding if a user is likely to change identity. Other reasons such as maturity of the adolescent, also plays a role in the likelihood of an identity change. To the best of our knowledge, we could not find any research which focused and explored changes to a unique attribute of a user on OSN i.e. username. We, in this work, addressed the research gaps, by conducting a study on Twitter users who change their profile attributes over time, and focused on a user's unique identifiable attribute -- username. 
\section{Limitations and Future Work}
We make the first attempt to study temporal changes in username attribute of a user and therefore acknowledge limitations of our work. Firstly, we monitor a small set of 10K users who have demonstrated the act of username change earlier in the past. We plan to expand this dataset, and generalize our findings. Secondly, we understand that our observation periods could have been longer. But with the frequency of username changes we observe, we think that inferences derived with our observation period of 4 months, can be generalized to a longer observation duration. Lastly, we receive limited user feedback to support our reasons and inferences. In future, we plan to perform an extensive user study and build automated systems to prove quantifiably that past identity change patterns could help in username suggestion, identity resolution and multiple identities correlation. We plan to investigate temporal changes to other profile attributes of Twitter users in our future work.

{
\bibliographystyle{IEEEtran}
\bibliography{references_icwsm}

\begin{thebibliography}{10}
\providecommand{\url}[1]{#1}
\csname url@samestyle\endcsname
\providecommand{\newblock}{\relax}
\providecommand{\bibinfo}[2]{#2}
\providecommand{\BIBentrySTDinterwordspacing}{\spaceskip=0pt\relax}
\providecommand{\BIBentryALTinterwordstretchfactor}{4}
\providecommand{\BIBentryALTinterwordspacing}{\spaceskip=\fontdimen2\font plus
\BIBentryALTinterwordstretchfactor\fontdimen3\font minus
  \fontdimen4\font\relax}
\providecommand{\BIBforeignlanguage}[2]{{%
\expandafter\ifx\csname l@#1\endcsname\relax
\typeout{** WARNING: IEEEtran.bst: No hyphenation pattern has been}%
\typeout{** loaded for the language `#1'. Using the pattern for}%
\typeout{** the default language instead.}%
\else
\language=\csname l@#1\endcsname
\fi
#2}}
\providecommand{\BIBdecl}{\relax}
\BIBdecl

\bibitem{Dewan:2013:MRM:2528228.2528235}
P.~Dewan, M.~Gupta, K.~Goyal, and P.~Kumaraguru, ``Multiosn: Realtime
  monitoring of real world events on multiple online social media,'' in
  \emph{I-CARE '13}.

\bibitem{ZafaraniL13}
R.~Zafarani and H.~Liu, ``Connecting users across social media sites: a
  behavioral-modeling approach,'' in \emph{KDD`13}.

\bibitem{perito}
D.~Perito, C.~Castelluccia, M.~A. K{\^a}afar, and P.~Manils, ``{How Unique and
  Traceable Are Usernames?}'' in \emph{PETS`11}.

\bibitem{ramsey2010brandjacking}
L.~Ramsey, ``Brandjacking on social networks: Trademark infringement by
  impersonation of markholders,'' \emph{Buffalo Law Review}, 2010.

\bibitem{curtin2010name}
T.~Curtin, ``The name game: Cybersquatting and trademark infringement on social
  media websites,'' \emph{Journal of Law and Policy}, 2010.

\bibitem{chhabra:phi.sh/ocial:-the-phishin:2011:yuqfj}
S.~Chhabra, A.~Aggarwal, F.~Benevenuto, and P.~Kumaraguru, ``{Phi.sh/\$oCiaL:
  The Phishing Landscape through Short URLs},'' \emph{CEAS`11}.

\bibitem{mcfarland2005motives}
D.~McFarland and H.~Pals, ``Motives and contexts of identity change: A case for
  network effects,'' \emph{Social Psychology Quarterly}, 2005.

\bibitem{jain2013seek}
P.~Jain, P.~Kumaraguru, and A.~Joshi, ``{@ I seek `fb. me': Identifying Users
  across Multiple Online Social Networks},'' in \emph{WWW`13 Companion}.

\bibitem{Irani}
D.~Irani, S.~Webb, K.~Li, and C.~Pu, ``{Large Online Social Footprints--An
  Emerging Threat},'' in \emph{CSE`09}.

\bibitem{Malhotra}
A.~Malhotra, L.~Totti, W.~Meira, P.~Kumaraguru, and V.~Almeida, ``{Studying
  User Footprints in Different Online Social Networks},'' \emph{{CSOSN`12}}.

\bibitem{goga2013exploiting}
O.~Goga, H.~Lei, S.~H.~K. Parthasarathi, G.~Friedland, R.~Sommer, and
  R.~Teixeira, ``Exploiting innocuous activity for correlating users across
  sites,'' in \emph{WWW'13}.

\bibitem{Xu:2012:MUP:2348283.2348358}
Z.~Xu, Y.~Zhang, Y.~Wu, and Q.~Yang, ``Modeling user posting behavior on social
  media,'' in \emph{SIGIR`12}.

\bibitem{abel2011analyzing}
F.~Abel, Q.~Gao, G.-J. Houben, and K.~Tao, ``Analyzing temporal dynamics in
  twitter profiles for personalized recommendations in the social web,''
  \emph{WebSci`11}.

\bibitem{fan2011incremental}
W.~Fan, J.~Li, J.~Luo, Z.~Tan, X.~Wang, and Y.~Wu, ``Incremental graph pattern
  matching,'' in \emph{SIGMOD`11}.

\bibitem{mislove2008growth}
A.~Mislove, H.~S. Koppula, K.~P. Gummadi, P.~Druschel, and B.~Bhattacharjee,
  ``Growth of the flickr social network,'' in \emph{WOSN`08}.

\bibitem{motamedicharacterizing}
R.~Motamedi, R.~Gonzalez, R.~Farahbakhsh, R.~Rejaie, A.~Cuevas, and R.~Cuevas,
  ``Characterizing group-level user behavior in major online social networks.''

\bibitem{5231867}
R.~Feizy, I.~Wakeman, and D.~Chalmers, ``Transformation of online
  representation through time,'' in \emph{ASONAM`09}.

\bibitem{burke2006identity}
P.~J. Burke, ``Identity change,'' \emph{Social Psychology Quarterly}, 2006.

\end{thebibliography}
}

\end{document}